\documentclass{article}

\usepackage{graphicx} 
\usepackage{hyperref}
\usepackage{url}
\usepackage{subfig}

\usepackage{multirow}
\usepackage{threeparttable}
\usepackage{bigstrut}

\usepackage{PRIMEarxiv}

\usepackage[utf8]{inputenc} 
\usepackage[T1]{fontenc}    
\usepackage{hyperref}       
\usepackage{url}            
\usepackage{booktabs}       
\usepackage{amsfonts}       
\usepackage{nicefrac}       
\usepackage{microtype}      
\usepackage{lipsum}
\usepackage{fancyhdr}       
\usepackage{graphicx}       
\graphicspath{{media/}}     

\title{A Small Leak Will Sink Many Ships: Vulnerabilities Related to Mini Programs Permissions
}

\author{
  Jianyi Zhang, Leixin Yang, Yuyang Han, Zhi Sun, Zixiao Xiang \\
  Beijing Electronic Science and Technology Institute \\
  Beijing\\
  zjy@besti.edu.cn}

\begin{document}
\maketitle

\begin{abstract}
As a new format of mobile application, mini programs, which function within a larger app and are built with HTML, CSS, and JavaScript web technology, have become the way to do almost everything in China. This paper presents our research on the permissions of mini programs. We conducted a systematic study on 9 popular mobile app ecosystems, which host over 7 million mini programs, and tested over 2,580 APIs to understand these emerging systems better. We extracted a common abstracted model for mini programs permission control and revealed six categories of potential security vulnerabilities in the permission environments. It is alarming that the current popular mobile app ecosystems (host apps) under study have at least one security vulnerability. We present the corresponding attack methods to dissect these potential weaknesses further to exploit the discovered vulnerabilities. To prove that the revealed vulnerabilities may cause severe consequences in real-world use, we show three kinds of attacks related to the mini programs' permissions. We have responsibly disclosed the newly discovered vulnerabilities, officially confirmed and revised. Finally, we put forward systematic suggestions to strengthen the standardization of mini programs.

\end{abstract}


\section{Introduction}

\begin{figure*}
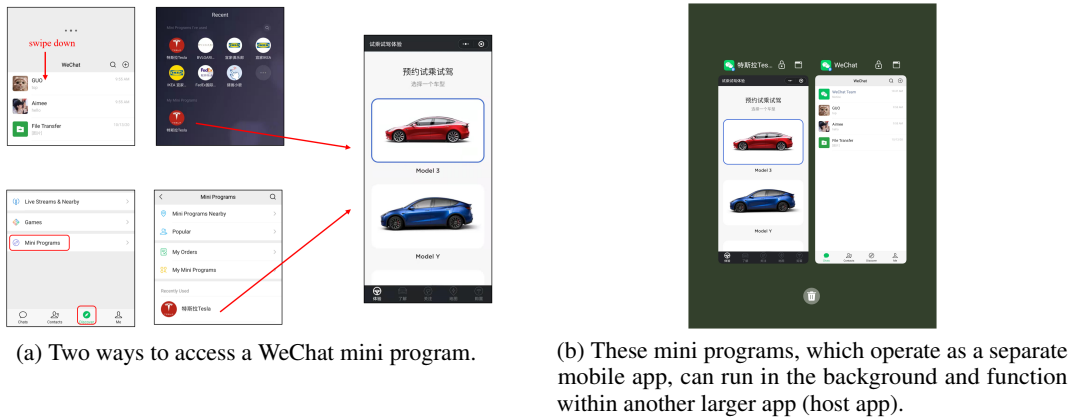

    \centering
    \subfloat[Two ways to access a WeChat mini program.]{
    \label{sub-fig-1}
    \begin{minipage}{0.4\textwidth}
    \centering
    \includegraphics[width=1.0\textwidth]{access_example.pdf}
    \end{minipage}
    }
    \qquad
    \subfloat[These mini programs, which operate as a separate mobile app, can run in the background and function within another larger app (host app).]
    {
    \label{sub-fig-2}
    \begin{minipage}{0.4\textwidth}
    \centering
    \includegraphics[width=0.5\textwidth]{run_in_the_background.pdf}
    \end{minipage}
    }
    \caption{Tesla mini program in WeChat. Source: Screenshot.}
    \label{fig:example}
\end{figure*}

Mini programs are light (commonly 2-4 MB) applications that run inside a specific mobile app (host app) \cite{Mini-Program-platforms-2021}. As a new form of mobile application, leveraging web technologies like HTML, CSS, and JavaScript, mini programs are taking over the iOS and Android app ecosystems in China. The mini program technology enables the "super app" to bundle features and capabilities into a single mobile native APP, which lets the users never need to leave this native app. We call this a host app. Many host app vendors, such as Tencent (WeChat), ByteDance (TikTok) and Alibaba (Alipay), provide their own framework to support the mini programs \cite{MiniApp-White-Paper}. There are various ways to launch a mini program in these host apps. Users can scan a QR code, directly search the name in the host app, share with a group or friend, launch it in a content article, or even link between mini programs. Figure \ref{sub-fig-1} shows how to launch the Tesla mini program in WeChat as an example. With the clear interface and fast loading times, mini programs are very easy to use. More and more people use mini programs in their everyday life and need not worry about installing too many apps. There is a mini program for just about anything in China, whether to pay bills, play games, order a taxi, or book a doctor's appointment. As of 2021, there are over 1.26 billion monthly active users (MAU) of WeChat \cite{WeChat-MAU}, 1 billion of TikTok \cite{TikTok-MAU}, and over 658 million of Alipay \cite{Alipay-MAU}. The total number of the mini programs' users is close to that of Facebook, the most popular social network worldwide, which has about 2.89 billion MAU \cite{Facebook-MAU}.

Since the widespread use of mini programs, any incorrect permission assignments or settings can result in serious security and privacy problems. However, there is not much research focus on this issue. That is not only because it is a new mobile application format, but also, more importantly, the permission structure of mini programs is entirely different from any other current permission-based security model. As we know, the mobile operating system (OS) is responsible for allowing or denying the use of specific resources at the app's run time \cite{barrera2010methodology}. The developers should declare a list of permissions that the user must accept before installing or running an application. Then the OS uses this security model to restrict every mobile app, native or hybrid, access to advanced or dangerous functionality on the device \cite{enck2009understanding}. Unlike the current access control models and methods, the permissions of mini programs are based on the host app authorization. That is, Android or iOS decides whether or not to allow the host apps to have some specific permissions, and the host apps authorize users to send pictures or other operations in the mini program. Therefore, from the view of OS, the permissions of mini programs and the host apps are the same, which means mini programs may apply for permissions from OS by using the reputation of host apps. It is impossible to control the permissions of mini programs directly. Once the host app does not authorize the correct permissions to the mini programs, it may cause security problems\cite{ma2019app}.

The permission issues in mobile applications have been fully studied in the past, and its permissions management mechanism is relatively complete and formal\cite{au2012pscout,chester2017m,rashidi2017android,reardon201950,liu2016follow,Android-vs-iOS-security,temporal2018sadeghi,demystifying2016backes,android-permissions-remystified2015wijesekera}. M-Perm\cite{chester2017m} can identify normal, dangerous, and third-party permissions requests in applications to detect permission abuse. DroidNet\cite{rashidi2017android} can provide advice on whether to accept or reject requests related to sensitive behaviors, helping users to implement low-risk resource access control on untrusted applications to protect user privacy. However, these studies cannot solve the relevant security issues in the permissions of the mini programs.

In this paper, we present our systematic analysis of the current mini programs’ permissions, where we dissect its framework, ecosystem, and potential vulnerabilities. We refer to the definition of sensitive permissions in Android and iOS and conduct a series of sensitive permission specification tests on mini programs. Specifically, we systematically studied 9 popular mobile app ecosystems, which host more than 7 million mini programs, and established an abstract model for the unique permissions application process of existing mini programs. According to the host App and OS allow or reject permission application, we divide the application process into three situations: Host App Allow, OS Allow; Host App Reject, OS Allow; OS Reject. The majority of the vulnerabilities belong to Host App Reject, OS Allow. According to this abstract model, we investigated more than 2580 APIs and revealed six categories of potential security vulnerabilities that are common in most of the mini programs we studied. According to the types of leaked information, we present three kinds of proof-of-concept attacks to analyze these potential weaknesses further. To prove that the exposed vulnerabilities may cause serious consequences on the real-world systems, we described three interesting cases of APIs and illegal mini programs. To mitigate the threat of these vulnerabilities, we have listed recommendations for mini programs, developers, and users.

Finally, to ensure that the mini programs' mechanism of different host apps had enough time to fix the vulnerabilities, we contacted them individually about the vulnerabilities several months before submitting this manuscript. This allowed several different host apps to finish patching the reported vulnerabilities in writing. We also disclosed the vulnerabilities to various security response platforms, including Tencent Security Response Center.

In summary, we have made the following contributions in this paper:

\begin{itemize}
    \item We deeply analyzed the current mini programs permission and presented a common abstracted model of the permission control of mini programs. To our best knowledge, we systematically studied the mini programs and the permissions for the first time.
    \item We have detected more than 2,580 APIs. Through large-scale tracking and analysis of sensitive permission APIs, we have found six categories of potential security vulnerabilities in the process of processing sensitive permission applications by mini programs.
    \item We conducted empirical research on the currently popular 9 host apps and revealed the security issues corresponding to the six types of potential security vulnerabilities we discovered in the real world. We also showed three real-life attacks on the mini programs permissions to prove that the revealed vulnerabilities may cause serious consequences in real-world use.
    \item Following the practice of responsible disclosure, we have reported all the discovered design flaws, officially confirmed and revised.
    \item  To mitigate these potential vulnerabilities, we put forward suggestions to strengthen the standardization of the entire mini programs' permission, thereby enhancing user privacy protection.
\end{itemize}

\section{Background}

\subsection{Framework of Mini Programs}\label{sec:Framework}

\begin{figure}
\centering
\includegraphics[width=0.5\linewidth]{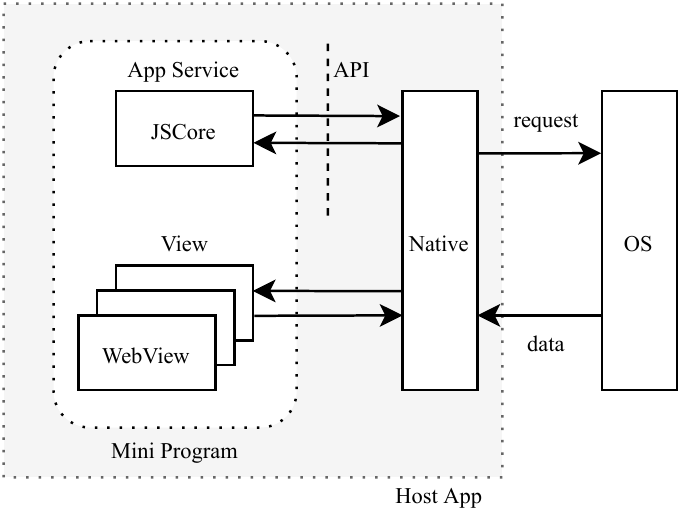}
\caption{The common framework of mini programs}
\label{fig:framework}
\end{figure}

Mini programs are a category of applications embedded in host apps without the need of downloading and installing~\cite{Analysis-of-the-development-of-WeChat-mini-program2018hao}. 
The flow of mini program framework consists of two components: View (the rendering layer) and App Service (the logic layer), which are respectively managed by two separate threads, as exhibited in Figure~\ref{fig:framework}. 
The interface of View is rendered by WebView component, which handles page displaying and the user event interaction behavior, while the App Service employs JsCore threads to run JavaScript, for controlling the generation and processing of mini programs data. 
The communication between two threads is relayed by the Native app (refers to the client). 


The host app in the framework determines whether the mini program has the permission for acquiring the specific data through the corresponding API. 
That is, OS determines whether to allow host apps to have some specific permissions, and host apps then transmit authorized data to the mini program through the API.
In other words, the permission of mini programs is inherited from the host app. 
Hence, if a host app does not properly manage the data and permission, data privacy and security issues will occur at its mini programs.
In this paper, we will focus on the App Service component to study the authorization mechanism of the host app and the design of API for mini programs. 


\subsection{Difference to Other Apps}
As a new emerging paradigm, mini program has the substantial difference to the existing native app, hybrid app, progressive web app (PWA), and instant app. 
Yet, there is little research on mini programs, especially for their potential security and privacy issues. 
We briefly summarize the difference between mini programs and other categories of apps here. 


A native app is a mobile application developed specifically for installing on device's OS like Android or iOS.
 A hybrid app looks like a native app, but at core, it is one kind of web app wrapped in a native container which loads 
 the information on the page when a user navigates through the application. 
 Although both hybrid apps and mini programs apply web technology, the hybrid app is still
 one type of native app that is under the management by OS. 
  In contrast, as a light version of the app, the mini program can only run on a native app’s (host app’s) interface and apply for the host app’s permissions. 

Similar to the mini program, PWA  also applies web technology, but differently, PWA is a type of webpage or website that runs in the browser and can be added to the home screen. 
Hence, the host environment of PWA is the browser and the OS manages the PWA's permission through the browser.
 Mini programs can be considered as one type of ``Instant'' app embedded in host apps for disposable interactions at a fraction of cost of an app. 
 The host environment is a platform with extra capabilities that can support seamless service and access control for the user data.
 
Google's instant app~\cite{What-is-instant-app} is very similar to mini programs. Both of them allow users to access the application's content without additionally installing the application, which can save the application space on the device. 
In essence, Google's instant app is still a native app and under the OS's permission control while the mini program is under the host apps'.



\begin{figure}
\centering
\includegraphics[width=0.5\linewidth]{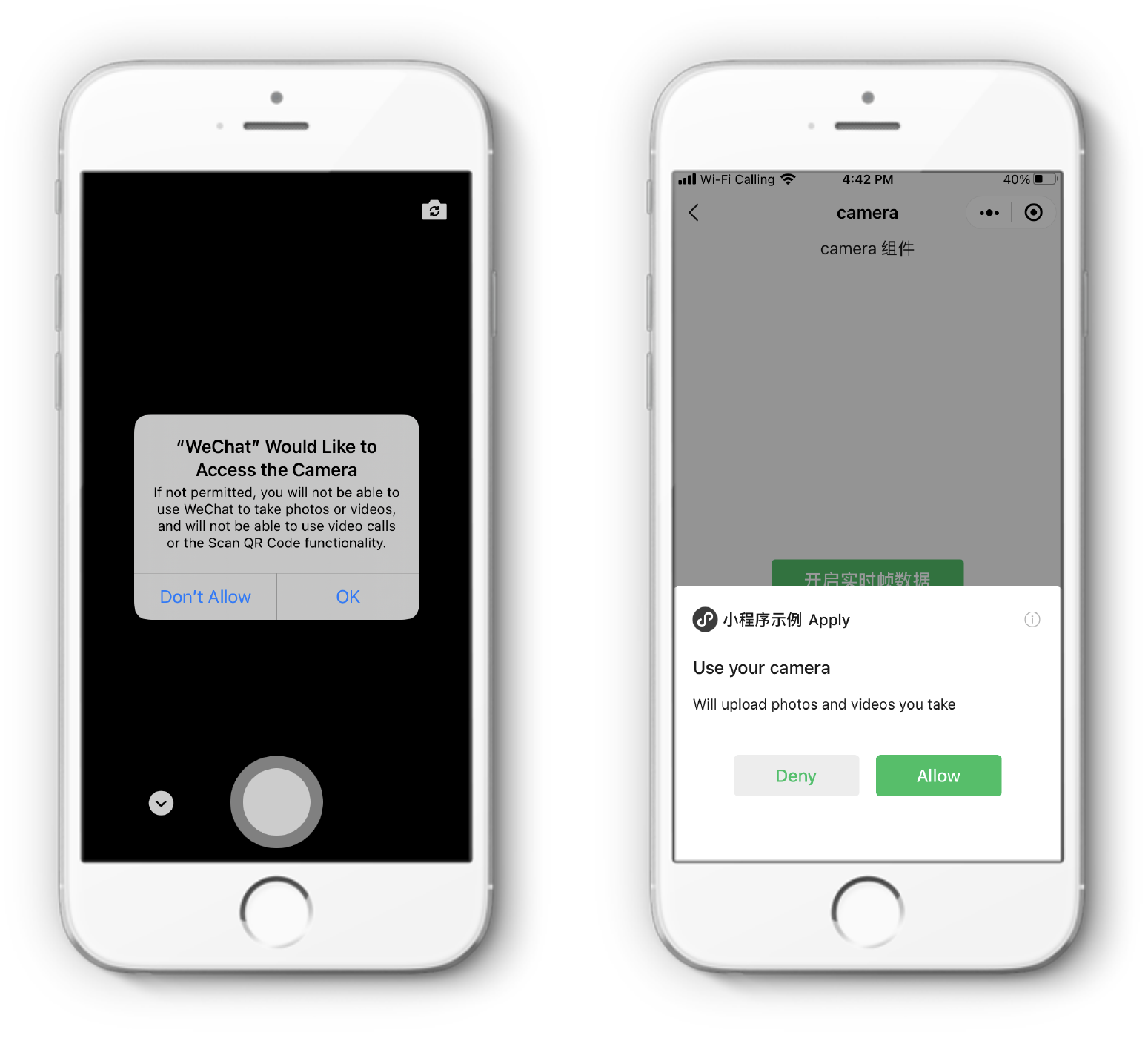}
\caption{Under the iOS system, the system permission pop-up prompt is displayed when Alipay requests runtime permission (left) and the pop-up prompt when a mini program in Alipay requests permission (right).}
\label{fig:runtime-permission}
\end{figure}

\subsection{Authorization} \label{subsec:auth}
Permissions in mobile apps can be divided into two types, install-time permissions and runtime permissions\cite{androidPermissions}. 
Here, the runtime permission is also called the dangerous permission, which is related to users' privacy and can access users' private data,
such as location information, contacts information, etc.
These information are considered to be sensitive and should acquire user's authorization for access. 
When requesting the runtime permission, 
the system will display a prompt window, as shown in Figure~\ref{fig:runtime-permission}. 

According to Section~\ref{sec:Framework}, the framework provides rich APIs to support the mini programs request a resource such as user information, location information, payment functions, etc. But the user does not authorize the API directly. In mini program development, the framework divided the dangerous APIs into multiple \textit{scope} according to the scope of usage. The users can select \textit{scope} to authorize. After a \textit{scope} is authorized, all of its APIs can be used directly.

\subsection{Permission Control Abstract Model} \label{subsec:MP Permission Control Model}

We conclude an abstract model common to all mini programs’ permission control, depicted as follows. 

\begin{figure}[ht]
\centering
\includegraphics[width=0.6\linewidth]{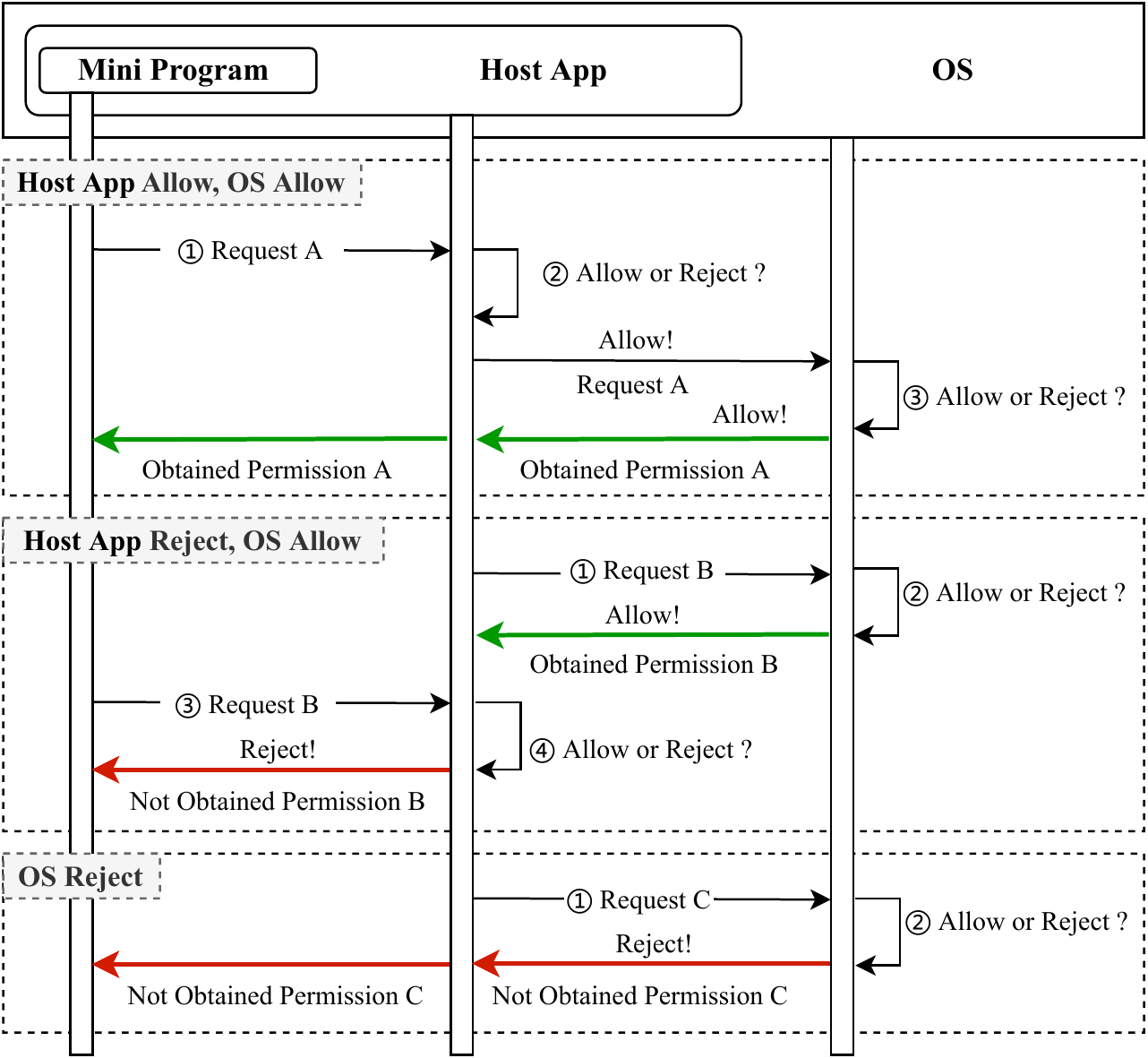}
\caption{The sensitive permission application process of mini programs. Mini programs run in specific mobile applications (host apps), while mobile applications run in the OS.}
\label{fig:application process}
\end{figure}

Mini programs are "sub-applications" built on mature mobile applications which are then built on the OS.
 So, mini programs need to pass two layers of barriers when 
 applying for sensitive permissions from users. 
That is, mini programs first apply to their host app where they are located and then the host App applies to the OS. 
As shown in Figure~\ref{fig:application process}, according to the host app and OS allow or reject permission application scenarios, we categorize the application process into the following three cases: 

\begin{itemize}

\item \textbf{Both Host App and OS Allow}
If a mini program applies for sensitive permission A, but the user has not 
applied this permission, the host app will pop up a window (as shown in the left of Figure~\ref{fig:runtime-permission}) to ask whether the user is willing to grant the permission. 
If it agrees, the host app will continue to apply for sensitive permission A to the OS, and the OS will pop up a window (the right side of Figure~\ref{fig:runtime-permission}) to ask if the user is willing to grant this permission to the mini program. 
The mini program will successfully obtain the requested sensitive permission once the user agrees. 
Later on, when the user uses the mini program again, he can directly call the interface.

\item \textbf{Host App Reject but OS Allow}
In this case, the host app has obtained permission B given by the OS, but when its mini program wants to apply for permission B, the user chooses to refuse and then directly enters the interface fail callback.
Hence, the mini program fails to obtain permission B. 
In our later analysis in Section~\ref{sec:Security Analysis}, most vulnerabilities we discover belong to this category.

\item \textbf{OS Reject}
If the host app does not get the permission C given by the OS, neither the host app nor the mini programs in it can get this permission.
It is worth noting that when the developer calls the API requiring authorization, the authorization setting will appear in the mini programs' authorization setting page regardless of the user's permission or not. 
The user can modify the authorization, and the calling result will change accordingly until the user actively deletes the mini programs. 
But user authorization setting is not synchronized with the server. 
If the user changes the device or switches his account on the same device, the authorization will be prompted again.

\end{itemize}

\section{Security Analysis} \label{sec:Security Analysis}

\begin{figure*}
\centering
\includegraphics[width=1.0\linewidth]{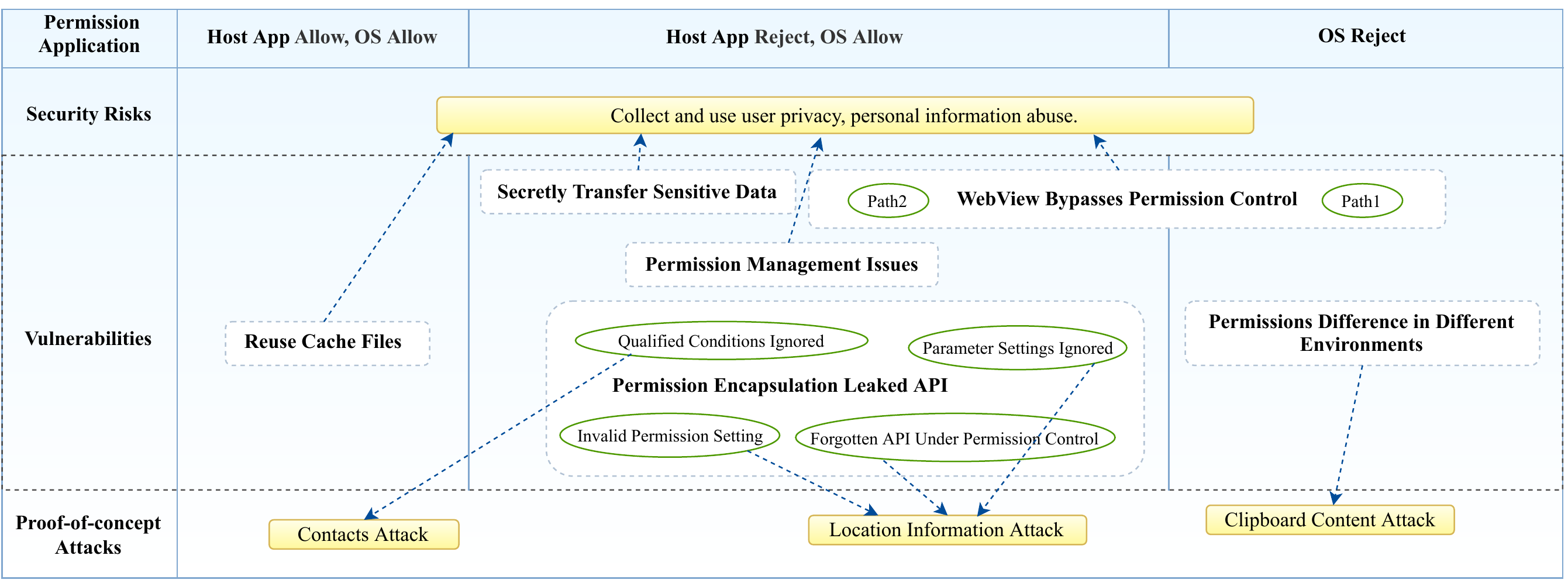}
\caption{The vulnerabilities and attacks in the mini programs permissions.}
\label{fig:Vulnerabilities and attacks}
\end{figure*}

After getting the background knowledge of mini programs, we start to perform the security analysis of mini programs' sensitive permission application process.
According to the three scenarios as we have discussed in Section \ref{subsec:MP Permission Control Model}, we discover six categories of potentially vulnerable modes during mini programs’ processing of sensitive permission applications, as presented in Figure~\ref{fig:Vulnerabilities and attacks}. 
Based on these vulnerabilities, we have created three proof-of-concept attacks. 
In this section, we will first illustrate the principle for determining the sensitive permission in Section~\ref{subsec:Regulation}.
Then, we elaborate the detailed vulnerability in Section~\ref{subsec:Potential vulnerabilities} and present our proof-of-concept attacks in Section~\ref{subsec:Proof-of-concept Attacks}.


\subsection{Whether to Request Authorization} \label{subsec:Regulation}

According to \ref{subsec:auth}, we know that the mini program framework puts the sensitive APIs into the \textit{scope}
. The framework determines whether the sensitive data can be obtained or processed by the mini program in the background as a rule for sensitive operations. For example, in the WeChat mini programs, \textit{wx.getLocation} obtains the current geographical location in the background. If there is no user authorization, the current location information will be transmitted without the user's awareness. Therefore, the mini program framework puts such APIs into \textit{scope} for unified authorization and management. In contrast, the API of \textit{wx.chooseImage} is to select pictures from local photo albums or take photos with cameras. Although the data involved are sensitive, the user needs to select and other interactive operations to send out the data. The framework will consider the user know and authorize this operation, and this data cannot be transmitted without the user's awareness. Hence, the mini program framework does not put this kind of API into \textit{scope}.




\subsection{Potential Vulnerabilities} \label{subsec:Potential vulnerabilities}

We conduct a large-scale analysis of APIs from different host apps and target APIs related to sensitive permission.
A total of 2,580 APIs are detected by us. 
Through large-scale tracking and analysis, examining the documentation, available source codes, and demos, we identify the following six categories of potential security vulnerabilities in the process of sensitive permission applications by mini programs.


\subsubsection{Reuse Cache Files}

When a user quits or deletes the mini programs, the cache files in the corresponding path should also be deleted, for reducing the user's storage space and avoiding being reused.
However, suppose the user closes the mini program in the host app and opens it again for operation. The previously cached temporary file is still there without being deleted. So, an attacker has the chance to obtain the local temporary file path of the previous file without the user's awareness.

\subsubsection{Permission Encapsulation in Leaked API} \label{subsub:3.2.2}

\begin{figure}
\centering
\includegraphics[width=0.6\linewidth]{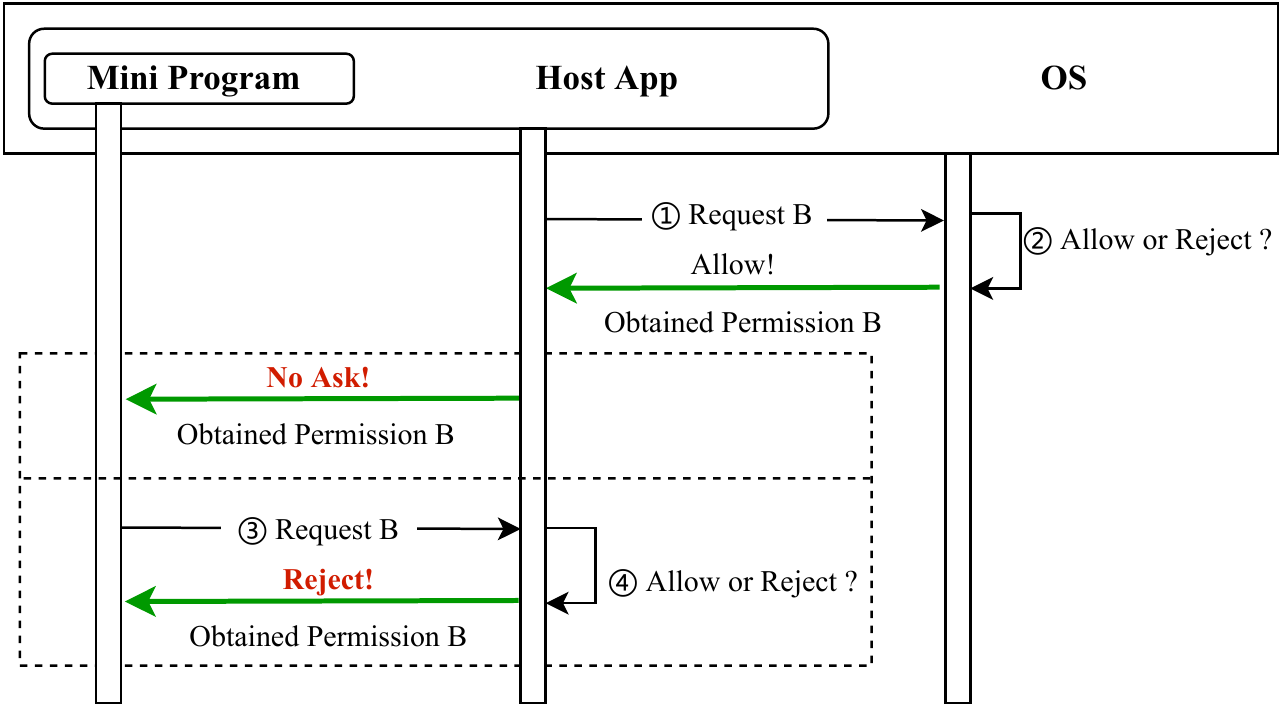}
\caption{When Host App Reject and OS Allow, mini programs illegally obtain the process of sensitive permissions.}
\label{fig:HA_Reject_OS_Allow}
\end{figure}

Some APIs related to sensitive permissions do not encapsulate permissions well (we refer to these APIs collectively as PEL-API). 
This means that any mini program can directly call these APIs to obtain the corresponding permissions, ignoring the need to apply permission within the host app. 
The application process of sensitive permissions corresponding to such an issue is shown in Figure~\ref{fig:HA_Reject_OS_Allow}. 
When the host app obtains a dangerous permission B from the OS, the mini programs can get this permission without asking the user's willingness in a pop-up window. 
Even having the pop-up inquiry and the user rejects it, the mini programs can still get this permission. 
In particular, we divide these vulnerabilities into the following four categories.

\noindent{\bf Qualified Conditions Ignored.} 
According to the description in Section~\ref{subsec:Regulation}, when the user must manually operate the sensitive permissions involved to view, select, or transmit private data, the mini programs will default that the user has allowed the use of rights in self-operation so that no pop-up will be made. 
In other words, there is no \textit{scope} (mentioned in \ref{subsec:auth}) corresponding to these APIs, and as long as the OS opens the relevant permissions for the host app, the mini programs can call these APIs and get data at will. 
At this time, sensitive data can only be used when the user interacts with mini program, which is equivalent to the user controlling his own sensitive information.
Mini programs cannot steal the user's sensitive information through this type of API. 
However, since the API of mini programs will be updated irregularly, the newly-launched API that involves the same permissions may have the function of stealing users' private data.
Many old APIs that involve the same permissions do not divide scope into corresponding ones, so these new APIs become a security risk. 
We believe that these APIs ignore qualification of sensitive permission (whether private data will be stolen without the user's awareness).


\noindent{\bf Forgotten API Under Permission Control.} \ 
The host apps handle whether the sensitive information can be passed to the mini program. For example, almost all host apps consider obtain ``location information'' is the dangerous permission. Whenever the mini program wants to use the current user's location, it will first apply to the user to ask for his consent. 
If the user refuses, the mini program itself cannot locate the user's specific location, regardless of whether the host app has obtained the permission. 
However, the designer neglected that some APIs should ask for authorization before sending the location information to the mini programs. 
In other words, these neglected APIs do not belong to any \textit{scope} (mentioned in \ref{subsec:auth}) and can be used without user authorization.


\noindent{\bf Parameter Settings Ignored.} \ 
From the point of view of its functions, some APIs may not actively obtain the user's sensitive permissions. 
However, the designer neglects the parameter settings in API, which will also steal users' privacy.
For example, \textit{my.chooseCity} in the Alipay mini programs is an API to open the city selection list. The parameter showLocatedCity indicates whether to display the currently located city. 
If it is set to true, the user's current city will be directly located regardless of whether the host program opens the location permission for the mini programs. 
If the user does not do anything, the background cannot see any content about the location information. 
However, as long as the user selects the area located by the system, the background will return to the current city and latitude and the user's longitude, even if the location permission is closed at this time.

\noindent{\bf Invalid Permission Setting.} \ 
The permission settings of some APIs are inconsistent with the descriptions in their official documents. To some extent, this shows that the API related to sensitive permissions in mini programs does not encapsulate permissions well. For example, \textit{wx.choosePoi} in the WeChat mini programs realizes the function of opening the map and selecting the location. It is indicated in the document that the invocation of this API requires the authorization of \textit{scope.userLocation}. However, in the actual test, we found that it is inconsistent with the official document description - the location can be selected without user authorization at all. When the user chooses precise positioning, the background will return the latitude and longitude data of the current user.

\subsubsection{Stealthily Transferring Sensitive Data}

\noindent{\textbf{Vertical}}: 
The application process for this vulnerability is shown in Figure~\ref{fig:HA_Reject_OS_Allow}. 
When some mini programs developed by serious cooperative companies involve sensitive permissions like geographical location, their host apps will ignore applying for permissions from users. The method shown as in Figure \ref{fig:HA_Reject_OS_Allow} can bypass user authorization for stealthily transmitting user sensitive information.

\noindent{\textbf{Horizontal}}: Different mini programs developed by the same company may share user information. 
All the user sensitive permissions acquired by the mini programs should be made public, and the user has the right to close the acquisition of sensitive permissions by the mini programs. 
Our empirical study found that some mini programs obtain and leverage the user information in their associated mini programs by default, including account information, shipping address, etc., but never 
request user's consent. 
This type of mini programs skips the permission application step, and the authorized permission in the settings is empty, resulting that a user is unable to close authorization of relevant user information. 
Users cannot fully control the transmission way and use the scope of their personal information, which risks continuous opening of permissions and unknown use of personal information.

\subsubsection{Permission Management Issues}

A mini program may continue to use sensitive permissions to collect users' privacy information even a user wants to turn off the permissions after using it. 
In particular, it can be divided into the following three situations.

\noindent{\bf Permission Setting Disappears.} \ 
The permission setting page provides convenience for users to manage permissions of mini programs. 
However, some mini programs may get permanent authorization after one-time authorization due to the lack of setting pages showing permissions.
In this case,  users cannot view what permissions they have granted to a mini program, and they cannot cancel the previously granted permission. 
As long as the host app is not uninstalled, the permission will remain open to the mini programs. 
Thus, a mini program can use the previous authorization to continuously gather and use the user's personal information, posing a security risk.

\noindent{\bf Permission Cannot be Deleted.} \ 
Regarding the validity period of authorization of mini programs, once a user explicitly agrees or rejects the authorization, such an authorization relationship will be recorded in the background until the user actively deletes the mini programs. 
However, the permissions some mini programs may not be able to delete, which can cause harmful consequences. 
These mini programs will be able to use the previous authorization to continue collecting and using users’ personal information.

\noindent{\bf Unable to Completely Remove Permissions.} \ 
Although the authorization settings of some mini programs are deleted after a user actively removes the mini programs, the sensitive personal information (such as ID number) involved will be retained. And such information can be leveraged for queries and other operations. 
At this time, the setting options of related permissions will not be in the mini programs' settings page, and the user cannot completely remove the permission.
Hence, although a user cancels the authorization, the related mini programs do not update the authorization in time, thus continuously collecting and using the user's personal information.
This leads to a risk of using a user's personal information for abuse.

\subsubsection{Webview Bypasses Permission Control} \label{subsub:webview bypass}


The mini programs can use the web-view page bearing component to open the H5 page in the mini programs. 
In this process,  the loaded H5 page needs to manually import JS files (which is a web development toolkit based on the host apps for web developers) provided by each platform. 
In this way, developers can apply the capabilities of mobile phone system such as taking pictures, selecting pictures, and location  with the help of host apps to provide users with a better web experience. 
After empirical study, we prove that webview can bypass the specified API when the mini programs or host apps do not prompt for permission application or the user refuses after prompting. 
There are two scenarios to illegally obtain sensitive permissions for mini programs after using the web-view component. 
First, the mini programs may completely ignore the OS's permission control over host apps and host apps' permission control over mini programs. 
They can directly access any sensitive permissions without notifying a user. 
Even when OS rejects sensitive permissions from a host app, the mini programs can still obtain such permissions through the web-view page-bearing component. 
Second, only the OS's permission control over host apps is considered, while the host apps' permission control over mini programs is ignored. 
The specific process is shown in Figure~\ref{fig:HA_Reject_OS_Allow}. 
In this case, if the OS's permission control on host apps is turned off, the mini programs will not obtain the corresponding sensitive permissions. 
If the OS's permission control on the host app is turned on, a user allows host apps to use certain sensitive permission. Hence, any mini program in host apps can obtain sensitive permission. 
We would like to remark that both scenarios may result in the disclosure of user privacy.

\subsubsection{Permission Issues in Different Environments}
The processing details for some APIs are not the same between different OS and versions. Since the mini program framework can not handle these APIs differently depending on the running environment, the same operation or program code will have different results.

For example, apps can read the clipboard contents without the user's manually selecting ``Paste'' when the user copies something. This is by design. Nevertheless, if the user copies sensitive information and leaves it on the clipboard, all apps can capture it and maybe send it to a remote server. Copying private things from a clipboard is risky. Different versions of OS have different feedback on this. The Android or old version of iOS will not inform the user when an application reads the clipboard. Many host apps will also be silent when the mini program reads the clipboard. If OS or host app does not consider the clipboard permission is dangerous, mini programs with access to the clipboard can steal the clipboard information of users in the background.

\subsection{Proof-of-concept Attacks} \label{subsec:Proof-of-concept Attacks}
The vulnerabilities mentioned above we found can result in users' sensitive information leakage. 
Here, we provide three proof-of-concept attacks to prove the seriousness of these vulnerabilities.

\subsubsection{Location Information Attack} 
An attacker can obtain the user's location information through the issues mentioned above discovered. 
For example, the second, third, and fourth issues discussed in Section~\ref{subsub:3.2.2} all involve location information leakage. 
An attacker may simply call an API with a bug in the design of host apps to obtain user's location information. 
Based on which, an attacker can infer personal information such as user's hobbies and sports patterns. 
More serious consequences can also be resulted, such as being tracked and personal attacks. 
If an attacker steals a massive amount of personal private information, the security and trust of the whole society will become an issue.

\subsubsection{Contacts Attack}
Some mini programs do separate the permission of ``contacts'', which can result in serious consequence. 
For example, the WeChat mini programs do not divide the permissions of ``contacts'' separately, so an attacker with ulterior motives can return and match similar mobile phone numbers in the background by calling \textit{wx.searchContacts}. 
Once a user is enticed to click the button bound to the event, a malicious mini program can traverse the user's address book information for fraud.

\subsubsection{Clipboard Content Attack} \label{subsubsec:clipboard}
Since the mini program framework does not restrict the apps to read the clipboard, developers only need a few lines of code to see what users have just copied. If users copy an online banking password to paste and leave this private information on the clipboard, a malicious mini program can read it in the background and see that data directly. The same goes for other sensitive data like names, addresses, credit card numbers, or even private photos. Mini programs can capture everything of user's clipboard and do whatever they want with it. The copied text could be sent to a remote server without user's awareness.

\section{Empirical Study}

This section presents our empirical study for analyzing the current mini programs permissions. 
Our goal is twofold. 
First, we collect the current popular mini program platforms (Table~\ref{tab:mini program list}), and then expose potential vulnerabilities as outlined in Section~\ref{sec:Security Analysis}.
We also exhibit the security issues exposed in the real world through detection. 
Second, we conduct case studies to show some real attacks on the permissions of mini programs, to prove that the revealed vulnerabilities may cause serious consequences in real-world use.




\begin{table}[]
\caption{The list of collected 9 host apps.}
\label{tab:mini program list}
\centering
\scalebox{0.84}{

\begin{threeparttable}
\begin{tabular}{lll}
    \toprule
\textbf{Company}           & \textbf{Host App} & \multicolumn{1}{l|}{\textbf{Monthly Active Users}} \\ \hline
\multirow{2}{*}{Tencent}   & WeChat                   & 1.26 billion \cite{WeChat-MAU}                                      \\ \cline{2-3} 
                           & Tencent QQ               & 595 million  \cite{QQ-MAU}                                      \\ \hline
Alibaba                    & Alipay                   & 658 million  \cite{Alipay-MAU}                                      \\ \hline
\multirow{3}{*}{ByteDance} & TouTiao                  & \multirow{2}{*}{400 million} \cite{TouTiao-MAU}                      \\ \cline{2-2}
                           & TouTiao speed Edition    &                                                    \\ \cline{2-3} 
                           & TikTok                   & 1 billion    \cite{TikTok-MAU}                                      \\ \hline
Baidu                      & Baidu                    & 607 million  \cite{Baidu-MAU}                                      \\ \hline
Multi Vendor               & QuickAPP                 & 130 million* \cite{QuickAPP-MAU}                                     \\ \hline
China UnionPay             & UnionPay                 & 10.6 million \cite{UnionPay-MAU}                                      \\ \hline
\end{tabular}
\begin{tablenotes}
       \footnotesize
       \item[*] Only contains Huawei's data.
\end{tablenotes}
\end{threeparttable}
}
\end{table}

\subsection{Mini Programs Ecosystem}

\begin{figure*}
\centering
\includegraphics[width=1.0\linewidth]{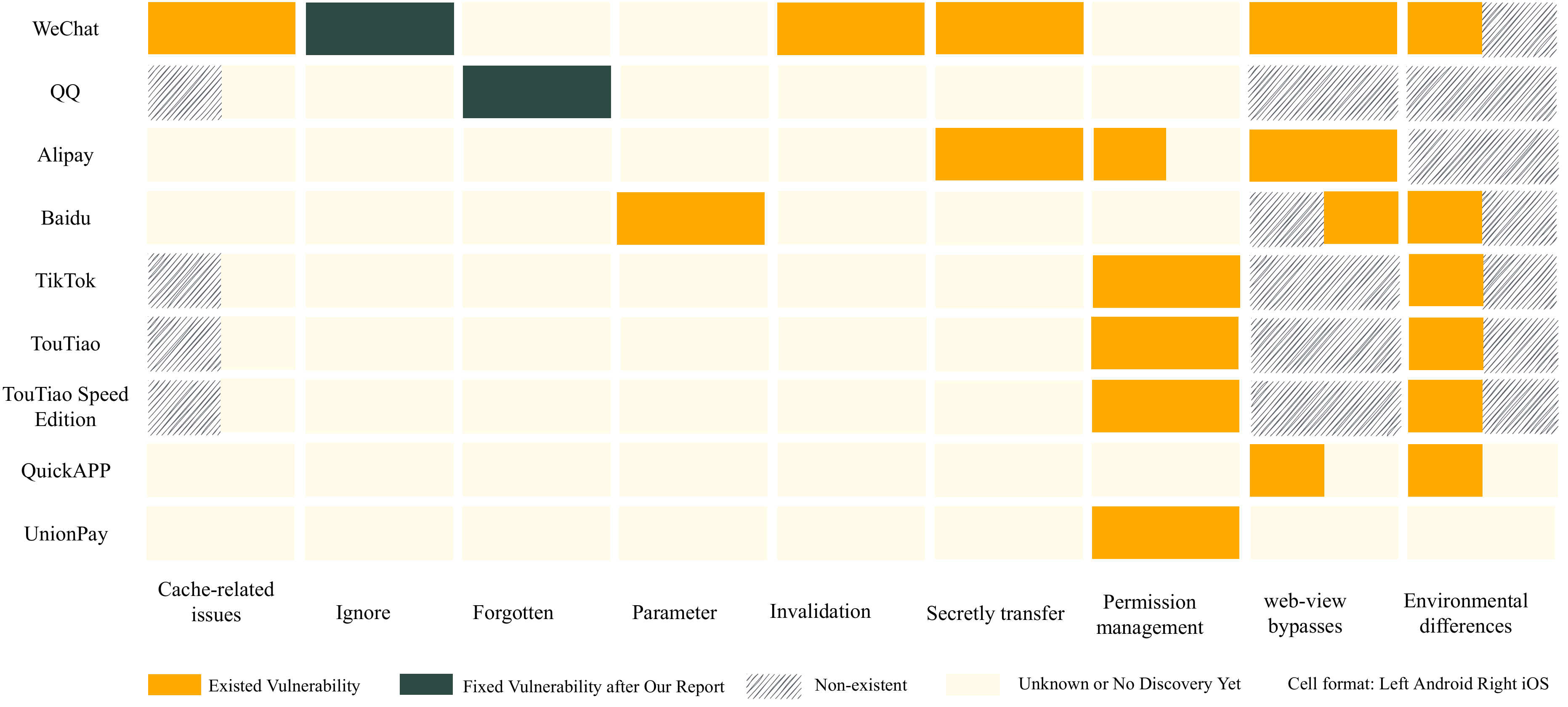}
\caption{The vulnerabilities distributions in the collected 9 host apps, where the vertical axis lists the names of host apps and the horizontal axis lists the vulnerabilities that we have discussed in Section 3. Cell format: $<$analysis result on Android$>$ | $<$analysis result on iOS$>$  }
\label{fig:vulnerabilities distributions}
\end{figure*}

As of June 2021, the number of mini programs in the whole network exceeded 7 million.
We have identified 9 popular host apps developed by 6 companies, which are listed in Table \ref{tab:mini program list}. 
These 9 host apps are used by tremendous amounts of users. 
Each host app has its own development tools. 
We apply the respective development tools to test on different host apps. 
Through empirical study, we discuss their vulnerabilities and  list them in Figure~\ref{fig:vulnerabilities distributions}.
In this figure, the orange blocks indicate that the host app has corresponding vulnerabilities, green blocks indicate that the host app has fixed vulnerabilities,  gray block indicates that the host app does not have such a vulnerability, and light yellow blocks indicate that it is uncertain if there exists such a vulnerability.

\subsection{Vulnerability Analysis}

\subsubsection{Vulnerable Caching Mechanism}
Our study discovered that in  WeChat, when a user closes the used mini programs (which have not been deleted from the recent use) and opens the mini programs again for operation, the previously cached temporary files still exist.
 Because the local temporary file path can be obtained in the background, the temporary file can be reused without the user's consent before being recycled. 
 In QQ and ByteDance, as long as the user exits the mini programs or reopens the previously used mini programs, the previous temporary files will be automatically deleted. Hence, they don't have the vulnerable caching issue. 
This vulnerability has not been found temporarily in mini programs of other host apps.


\subsubsection{Vulnerable Issues API}




\begin{table}[]
\caption{The list of Collected PEL-APIs.}
\label{tab:Vulnerable problem API}
\centering

\scalebox{0.9}{
\begin{tabular}{lll}
\hline
\textbf{Vulnerabilities}   & \textbf{Host App}   & \textbf{API}                 \\ \hline
Ignore                     & WeChat              & \textit{wx.searchContacts}            \\ \hline
\multirow{2}{*}{Forgotten} & \multirow{2}{*}{QQ} & \textit{MapContext.moveToLocation}    \\ \cline{3-3} 
                           &                     & \textit{MapContext.getCenterLocation} \\ \hline
Parameter                  & Alipay              & \textit{my.chooseCity}                \\ \hline
Invalidation               & WeChart             & \textit{wx.choosePoi}                 \\ \hline
\end{tabular}
}

\end{table}

As discussed in Section~\ref{subsub:3.2.2}, this kind of issue comes from that the API related to sensitive permissions does not encapsulate permissions well. 
In order to find APIs with such vulnerabilities, we set two criteria for analysis: 
1) whether Android and iOS treat the involved permissions as permissions that need to be granted by the user; 
2) whether using permissions by these APIs involves the interaction between users and mini programs. 
If the user must manually operate to view, select or transmit private data, then the mini programs default that the user has allowed the use of the permissions without prompting, so these APIs are not within the scope of our research. 
Table~\ref{tab:Vulnerable problem API} summarizes the APIs that we have found so far that are related to sensitive permissions but do not encapsulate permissions well.

\subsubsection{Vulnerable Permissions Transfer}

\noindent{\textbf{Vertical}:} "Amap" in the Alipay mini program can be opened directly to accurately locate the user, ignoring the application of the mini program to the user for location permission. 
Although Alipay and AMap have reached an in-depth cooperative relationship (both belong to Alibaba Group), this does not mean that their operations can bypass the user's willingness. 
No abnormality was found in the public test code of ``demo'' provided by Alipay, but using this mini program alone can directly locate users accurately. This shows that there may be an inconsistency between the source code of ``demo'' and the public test code provided. They may use other ways to bypass the user authorization to transmit the user's location secretly.

\noindent{\textbf{Horizontal}:} Some companies may share user information among different mini programs. For example, after logging in the ``Pinduodu'' mini program in WeChat, the same login information is displayed directly when the user first uses another mini program named ``Pinduoduo Coupon'', which is from the same company Pinduoduo Inc. This type of mini program skips the permission application step, and the authorized permission in the settings is empty. Hence, the user cannot close the authorization of relevant user information. Users cannot fully control the dissemination of their personal information and the scope of use, which will lead to the continuous opening of permission and the risk of personal information being used in unknown circumstances.

\subsubsection{Insecure Permission Management}

The permission management of UnionPay mini programs is rather messy. 
Since the security of UnionPay is mainly related to user identity information and payment, it is not very strict to consider other sensitive permissions such as microphones, geographic locations, cameras, and photo albums. On the privacy settings page of the UnionPay APP, we can see the authorization information for mini programs. 
However, only the authorization of the phone number and identity information (name, ID number) is displayed. In contrast, other permissions such as location, access to mobile phone albums, and the camera are ignored, and there is no permission setting page inside mini programs. 
When a user grants mini programs sensitive permission such as a microphone, geographical location, camera, and photo album, as long as the UnionPay app is not uninstalled, the mini programs can always access these permissions.

We found that, in  ByteDance, the permission of mini programs cannot be deleted with the deletion of the mini programs.
 In the official ByteDance document, there is no clear explanation on the validity period of the authorization and whether the authorization will remain open after deleting mini programs. 
 Hence, mini programs under ByteDance may use previous authorizations to continuously access and collect user personal information, posing a security risk.

Although the authorized information of Alipay mini programs can be deleted after a user actively deletes the mini program,
 the sensitive personal information involved before (such as ID number) will be retained by individual mini programs. 
We have canceled the authorization and removed the mini program, but the relevant information can still be consulted when entering the mini programs again after a certain period. At this time, the permission setting of the mini program becomes empty, and the user cannot set the permission on the setting page. This situation indicates that after the user cancels the authorization, the related mini programs do not update the authorization in time and may continue to collect and use the user's sensitive information after the user has canceled the authorization.

\subsubsection{Insecure Webview Component}

Some mini programs can carry web pages through the webview component. The communication between the webpage and mini programs is realized by the interfaces provided by the webpage development kit based on host apps provided by each platform for web page developers. None of the communication outside the provided interface is supported. Neither QQ nor ByteDance found any interfaces involving dangerous permissions. In the WeChat host app under Android, it can be determined that the user's storage information can be obtained using the webview component. That is, the first scenario described in Section~\ref{subsub:webview bypass} can be used to bypass the normative permission control. The difference is that in WeChat mini programs under iOS, it is used to obtain the user's album permission, and the second scenario is used to obtain the user's camera permission. 
The mini programs in Alipay bypass the standard permission control through the second scenario described in Section \ref{subsub:webview bypass}, to obtain sensitive permissions such as the user's camera, photo album, location, etc., even if these permissions are turned off in the setting page of mini programs at this time. 
That is, when users reject these sensitive permissions, the sensitive information can still be obtained by using the webview component. 
The mini programs under Baidu, in the iOS environment, can use the second scenario to obtain the camera permission to take pictures. 
QuickApp in Huawei can obtain the users' stored information (pictures, audios, videos, documents, etc.) through it. This will cause users to disclose their precise location to the attacker without their awareness.
If the user's mobile phone turns on the GPS when taking photos, then once the mini programs can access the camera and photo album, the malicious mini programs only need to filter out which photo taken by the user's mobile phone to know its geographical location.


\subsubsection{Vulnerable Clipboard Mechanism}

\begin{figure}
\centering
\includegraphics[width=0.6\linewidth]{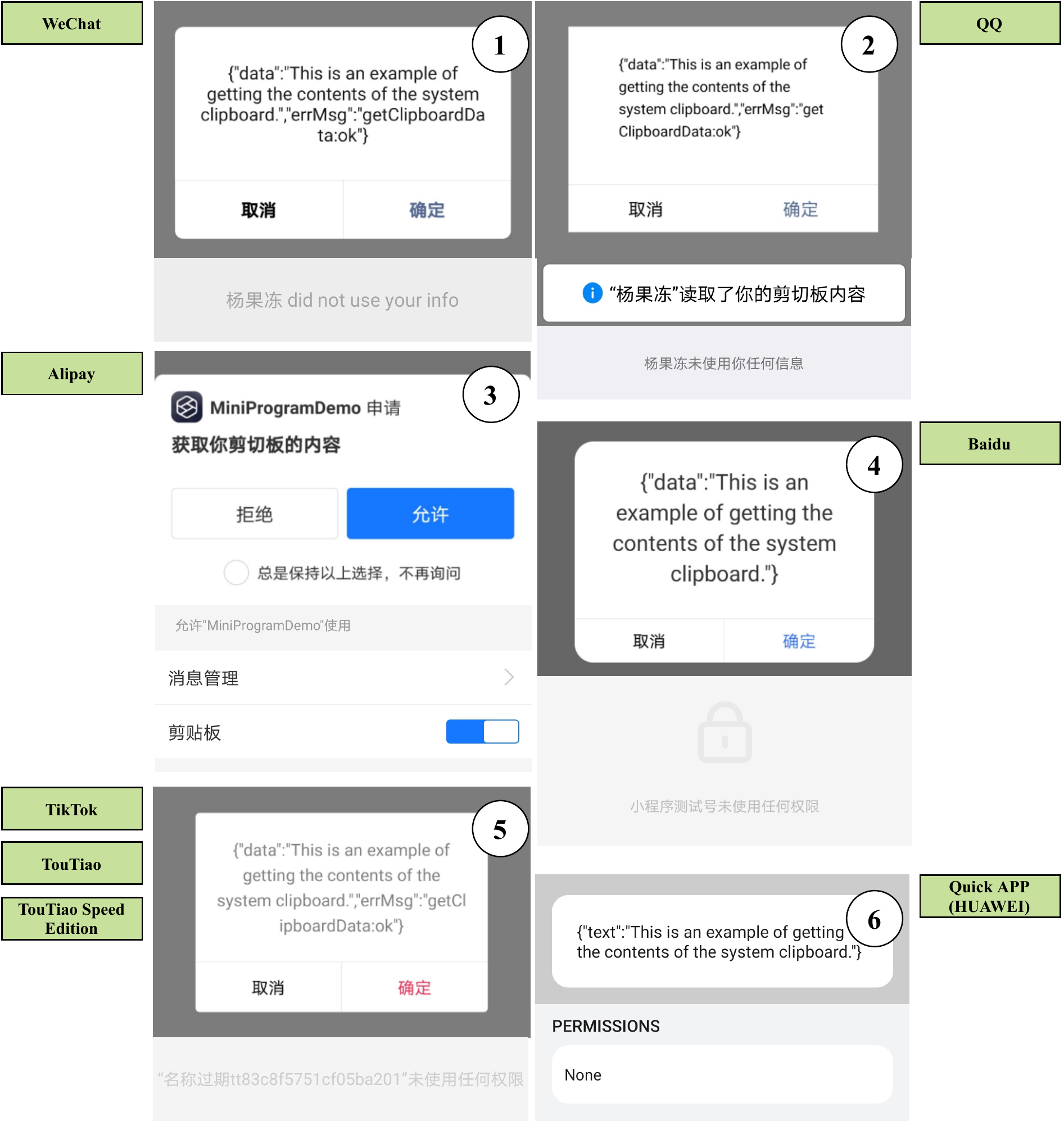}
\caption{Examples of getting clipboard contents in Android. Only case 2 and case 3 will prompt the user when the mini programs get the contents of the user's clipboard.}
\label{fig:getting clipboard contents}
\end{figure}

After iOS 14 upgrade, there will be a pop-up to informs the user when an app reads the contents of the clipboard. Hence, there will be a notification when the mini programs read the clipboard no matter which host apps they run inside. 

Android is more complex since there are multiple versions. Xiaomi MIUI12(MIUI is a third-party mobile phone OS deeply optimized, customized, and developed by Xiaomi based on Android) splits the permission of the clipboard independently, and users can monitor the reading and writing behavior of each application. However, in other Android phones (such as Huawei, Vivo, etc.), the permissions control of the clipboard will not inform the users. Under these OS versions, we test the clipboard permissions on different host apps in the Android environment, and check their corresponding prompts, with results shown in Figure~\ref{fig:getting clipboard contents}. 
In order to exhibit that the mini programs have obtained the contents of a user's clipboard, we display a modal dialog box. We found that 6 out of the 8 host apps (cases 1, 4, 5, 6 in Figure \ref{fig:getting clipboard contents}) did not give any prompt to the user when obtaining the clipboard information, which may be vulnerable to the theft of clipboard content described in Section~\ref{subsubsec:clipboard}. 
Cases 2 and 3 in Figure~\ref{fig:getting clipboard contents} show a design example that the mini programs will prompt when obtaining the contents of a user's clipboard. 
However, case 2 only reminds the user that the mini programs have obtained the clipboard's contents through a bubble, but the user only knows that it cannot be blocked (same as the prompt in the iOS environment). In case 3, obtaining the clipboard's content is set as the user's permission to operate. 
The user must select ``Allow'' before the mini program can obtain the corresponding information. We consider these designs with the best security usability practice.

It should be noted that, although it is set as the user's permission to operate in case 3, there is still an issue with Alipay. In version 10.2.26.8000, Alipay will pop up a window to ask the user "Apply to obtain the contents of your clipboard", and the contents of the clipboard have been pasted on the page before clicking "Reject" or "Allow" (this is equivalent to the mini program still being able to obtain the clipboard information after the user refuses authorization). However, in version 10.2.23.7100, users cannot get any content after clicking Reject.

\subsection{An Example} \label{appendix:Experimental Evaluation}

\begin{figure*}
\centering
\includegraphics[width=1.0\linewidth]{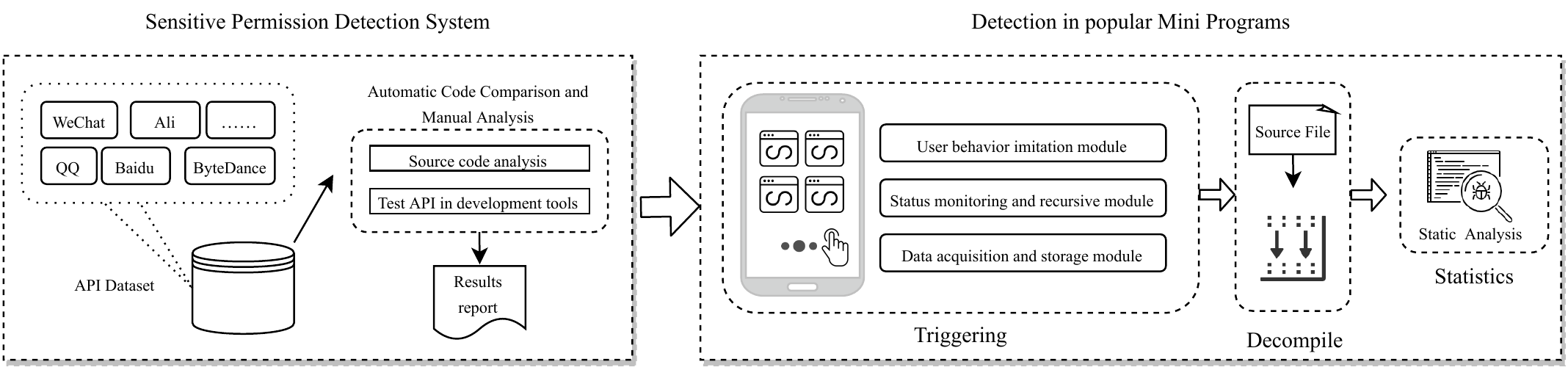}
\caption{Overview of measurement methodology}
\label{fig:methodology}
\end{figure*}

WeChat is an application in China with more than 1 billion active users, and it is also one of five applications in the world that have surpassed this milestone. 78\% of Chinese people aged 16-64 are using WeChat~\cite{WeChat-Revenue-and-Usage-Statistics}. So, we take WeChat as an example and detect the API with permission issues in the popular mini programs and expose security issues hidden in the real world. 
MiniCrawler\cite{zhang2021measurement}, an open source WeChat mini program crawler, can be used to automatically download, unpack and index the mini programs from WeChat. However, the tool has strict requirements on the WeChat version, and the latest WeChat version API has been replaced, so the tool cannot be used.
Our measurement method is shown in Figure~\ref{fig:methodology}.


\subsubsection{Implementation}

In order to ensure the smooth progress of the experiment, we did not choose the Android simulator (according to the previous test experience, running small programs with the Android simulator will be very stuck). 
Instead, we select an Android entity machine with ROOT permission to carry out our operation, as shown in Figure~\ref{fig:methodology}. 
Our experiment is mainly divided into the following three parts:

\paragraph{Triggering}
We use Airtest, a UI automated testing tool based on image recognition and poco control recognition, to achieve automated clicks on mini programs to obtain the source files (that is, downloading the source file package of mini programs from the server to local). 
Notably,  since mini programs can be used in parallel with the host apps after clicking, each mini program needs to be completely exited after the simulation click is over. Otherwise, too many open mini programs will cause the Android emulator slow. In the process of implementation, to achieve as far as possible that each page of mini programs can be loaded successfully to obtain all the source files of mini programs, we adopted LSH image detection algorithm for near repetition to determine whether the current page is consistent with the previous page if it is consistent, and treat them as the same page. If it is inconsistent, it is considered as the next page. 
In order to illustrate our experiment more clearly, the simulation click part is separated into the following three sub-modules:

\noindent{\textbf{User Behavior Imitation Module:}} \ 
The function of this module is to imitate the process of users clicking into mini programs. The program completely replaces the human operation: Automatically clicking the ``search'' icon to enter the search bar through the simulated click technology, automatically input the name of mini programs stored in the sample library in the search bar through the file retrieval technology, and identify the accurately searched mini programs based on the image recognition and poco control, to imitate the user behavior and enter mini programs.

\noindent{\textbf{Status Monitoring and Recursive Module:}} \ 
The main function of this module is to monitor the current interface position of mini programs in real-time and recursively scan the page content. By defining a page scanner and combining LSH image detection algorithm, the depth of the current page is monitored in real-time to determine whether to enter the next interface or call back to the previous interface through recursive function, for scanning as many different interfaces of the same mini programs as possible and trying to get all the cache packets of mini programs.

\noindent{\textbf{Data Acquisition and Storage Module:}} \ 
The main function of this module is to obtain the source files of mini programs. Take WeChat mini programs as an example. In the scenarios of WeChat version 8.0.2, the source files of a certain WeChat mini program are all in the path ``/data/data/com.tencent.mm/MicroMsg/.../appbrand/pkg''.
 Then, we use the shell command to dump the source code of the mini program to the folder named after the WeChat mini program to facilitate the follow-up work. At the same time, it is also necessary to delete the code packages that have been transferred to other locations in a specific folder in time to prevent confusion with the code packages of the next WeChat mini program.

\noindent{\bf Decompile:} \ 
We obtain the source code of the specified mini programs through decompilation. 
We have added a graphical operation interface to wxappUnpacker \cite{wxappUnpacker} for the decompilation script and configured Nodejs, the running environment required for the decompilation script, which can decompile the mini program source files obtained in the previous step to obtain their source code.

\noindent{\bf Statistics:} \ 
This step counts the usage of APIs with permission issues in popular mini programs in the world.
 Based on the issues found in section \ref{subsec:Potential vulnerabilities}, we sorted out different mini programs platforms and listed APIs that have permission issues. Screen these APIs from the source code of the mini programs, and measure the usage of the problematic API in popular mini programs in the real world.

\subsubsection{Results}\label{subsub:Results}

\begin{figure}
\centering
\includegraphics[width=0.6\linewidth]{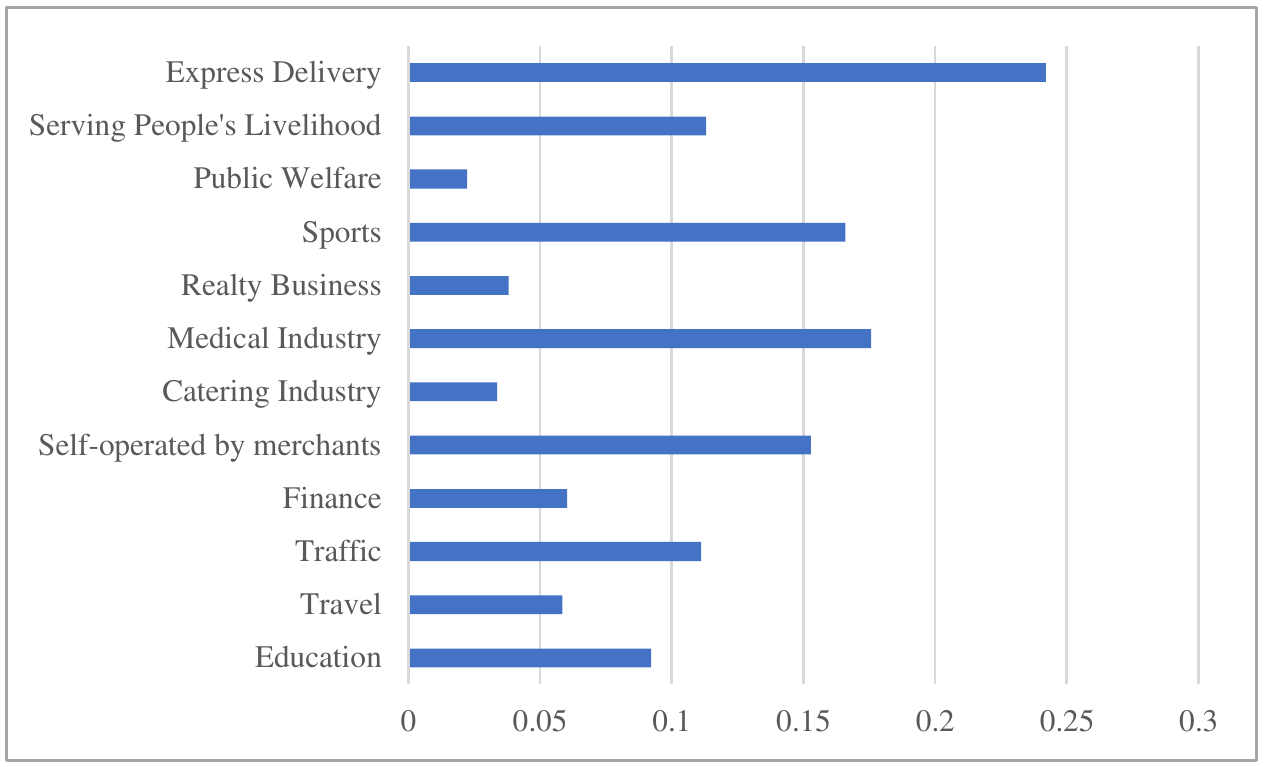}
\caption{The proportion of \textit{wx.getClipboardData} in different categories of WeChat mini programs.}
\label{fig:getClipboardData}
\end{figure}

We refer to the "Aladdin Index" (an intelligent platform for data analysis of the entire network of mini programs), to select 578 popular mini programs for research, and get 977 mini programs packages after decompression. We take the permission of the clipboard as an example to detect the severity. Figure \ref{fig:getClipboardData} shows the proportion of \textit{wx.getClipboardData} in various WeChat mini programs. Almost every kind of WeChat mini program use the vulnerable API as a standard API, which has the potential risk that some mini programs apply these API to invade user privacy.

\subsection{Case Studies}
In this section, we present our case study of some representative APIs and illegal mini programs.

\subsubsection{Stealing Location Information}
In the QQ mini program development tool, the index.js file used \textit{qq.createMapContext} to create a \textit{MapContext} object; then we used \textit{MapContext.moveToLocation} to move the map center to the current location (at this time, the map component show-location is set to true); then use \textit{MapContext.getCenterLocation} to get the latitude and longitude of the current map center. In the whole process, the user's geographic location, latitude, and longitude can be accurately obtained in the background without the user's authorization. When working with Tencent's location service to accurately obtain the location, the user's specific location can be obtained. We confirmed that the QQ mini programs could obtain the coordinates of the center point of the current location without the user's authorization, regardless of the Android or iOS system, and successfully transfer the coordinate value, specific location, and other information to the mini program background. Without authorization, the mini programs can obtain the user's precise location. Once the mini programs associate the location information with the account information in the background, the user's personal information is completely exposed. This is equivalent to unauthorized access to the user's whereabouts (belonging to personally sensitive information), and the mini programs are suspected of obtaining personal information in violation of regulations.

\subsubsection{Stealing Contacts}
The API provided by the WeChat mini programs: \textit{wx.searchContacts} is to find the contacts and match a similar mobile phone number. This API does not specify the number of calls within a period. Suppose the \textit{phoneNumber} parameter (the number to be searched) in this function is written in a loop. In that case, most of the information in the user contacts can be obtained by traversing in sequence. Specifically, set a button in index.wxml to bind an event. Write the API \textit{wx.searchContacts} into this event in the index.js file, and click this button, the foreground will not respond, and the background will get the information of the user's contacts. At this time, if the host environment WeChat has already obtained the contacts' permission, the mini programs can obtain part of the contact information without the user's awareness. After testing, we have confirmed that WeChat mini programs can use this method to obtain information of some users' contacts and successfully pass the obtained information into the mini programs' background, whether in Android or iOS, without the user's authorization and the user's awareness. Attackers can take advantage of this vulnerability and bind \textit{wx.searchContacts} to an inductive button to entice users to click. Sensitive information in the user's address book may be read and uploaded to the background, resulting in information leakage.

\subsubsection{Stealing Clipboard Information}
Take the WeChat mini programs as an example. During the test, we write \textit{wx.getClipboardData} in the onload function in the JS code so that the clipboard content can be easily obtained without the user's awareness. Even if it is not written in the onload function, binding this event to the button control can also trigger it. In the real world, mini programs can write some inductive slogans on the button to induce the user to click and then obtain the contents of the user's clipboard. After a large number of tests, we have confirmed that many host apps (WeChat, ByteDance, Baidu, QuickAPP) can obtain user clipboard information and successfully transfer the obtained information into the mini programs' background without the user's authorization and awareness. Suppose the copied content is not destroyed after the user pastes it into an application. In that case, the content can still be obtained when the user opens a mini program, thus causing the leakage of sensitive information of the user. For example, when a user copies the name of a certain product, after opening a shopping mini program, it can read the content of the user's clipboard and upload the private information in the background. The vendor of this mini program will know the user may want to buy this product and push similar commodities or analyze the user's behavior to push advertisements accurately.

\subsection{Responsible Disclosure}

We reported the results of our investigation to the Tencent Security Response Center on May 6, 2021, and August 27, 2021, and received their vulnerability confirmation. We also reported vulnerabilities regarding location and contacts permissions to CVE, received their confirmation of the vulnerability on May 18, 2021, and August 29, 2021, and obtained CVE numbers CVE-2021-33057 and CVE-2021-40180.


\section{Discussion} \label{sec:discussion}

\subsection{Limitations and Future Works} \label{subsec:Limitation}

1) We conducted a series of mini programs' permissions tests based on personal accounts. In other words, the tested APIs are all for individuals. According to our statistics, the number of APIs open to non-individual developers and certified mini programs is 4.6\% of the total API numbers.
This type of interface usually includes obtaining the user's mobile phone number, sports data, etc. Different host apps have different attitudes towards such interfaces. For example, the WeChat mini programs allow individual developers to obtain user sports data, while in the Alipay mini programs, this interface is only open to corporate users. Generally, those interfaces that are only open to enterprise users will be more strictly managed by host apps since these can obtain more data. It is very difficult for us to test the APIs used by the enterprise accounts. In the future, we will cooperate with enterprises to test and research these APIs.

2) In the \ref{appendix:Experimental Evaluation}, we take the permissions of the clipboard in the WeChat mini programs as an example to test it in the mini programs that are currently online. The percentage of detected problems is not large. However, it does not mean that the vulnerabilities we put forward do not appear in mini programs in other host apps. As more global companies join the mini programs related discussion, the number of emerging host apps supporting mini programs (such as UnionPay, NetEase) increases. They just appeared, and the vulnerability proposed in this article may not exist. When host apps gradually add new mini programs' functions or APIs, they may also make the vulnerabilities we found in this article, so the host apps still need to pay more attention to the permissions in the future.



\subsection{Mitigation Measures}

The purpose of our empirical analysis was to draw people's attention to the neglected security issue of the improper use of sensitive permissions in mini programs. In the face of the rapid development of mini programs and the sharp rise in the demand for personal privacy protection of mini programs, it is necessary to strengthen the coordination of host apps, apps vendors, and users to form a management system for personal privacy protection of mini programs. 

The responsibility of host apps is the biggest. In essence, mini programs are still software that provides various services for mobile users. Mini programs are put on shelves through the host apps, and host apps should strictly control the management of permissions, actively rectify APIs related to users' sensitive permissions, and protect user privacy with the most significant effort. For the apps vendors, it is necessary to consider how the existing OS handles specific APIs and the results feedback. For example, face the clipboard issues, the developer can highlight any word in the app and select ``Copy'' or automatically clear the clipboard after a while. Users also need to enhance the security awareness to protect personal information when using mini programs, be vigilant against mini programs of unknown origin, and do not quickly authorize their private information to the mini programs to avoid being illegally collected and leaked, causing unnecessary losses. At the same time, users should be encouraged to report illegal activities actively.

\section{Related Work}

\paragraph{Reason Exploration.} 

Almomani et al.\cite{almomani2020comprehensive} demonstrated, discussed and compared the latest technologies in the field of Android permissions, and conducted the latest research on Android permissions, revealing that the Android permissions faces various security issues. Zheran Fang et al. \cite{Permission-based-Android-security2014fang} investigate the arising issues in Android security, including coarse granularity of permissions, incompetent permission administration, insufficient permission documentation, over-claim of permissions, permission escalation attack, and TOCTOU (Time of Check to Time of Use) attack and put forward several methods to further reduce Android security risks. Joel Reardon et al.\cite{reardon201950} searched for sensitive data being sent over the network for which the sending app did not have permissions to access it by mechanisms to monitor the application’s runtime behavior and network traffic. They found that apps can circumvent the permission model and gain access to protected data without user consent by using both covert and side channels and determined how this unauthorized access occurs. Mujahid \cite{studying2018mujahid} implement a technique in a tool called PERMLYZER, which automatically detects permission issues from apps APK.

\paragraph{Detection and Protection.}

In the Android, the most common is the permission management mechanism of Android\cite{alepis2019unravelling,rethinking2016zhang,taxonomy2016sadeghi,Securing-android2015tan2015}. DroidNet\cite{rashidi2017android} is an Android permission control and recommendation system, which is an Android permission control framework based on crowdsourcing. It provides recommendations on whether to accept or reject the permission requests based on decisions from peer expert users, which can help users implement low-risk resource access control for untrusted applications and protect users' privacy. HybridGuard\cite{phung2017hybridguard}, a framework based on the subject authority and fine-grained policy execution for web mobile applications, can accurately monitor all web codes to ensure the security of mobile applications, in which an interception and policy code is implemented in a single JavaScript file, and whether to intercept them is determined by wrapping API about device resource access and DOM operation and checking the policy. M-Perm\cite{chester2017m} is a detection tool that combines string analysis and static analysis to identify normal, dangerous, and third-party permission requests in applications to detect permission abuse. Cusper\cite{tuncay2018resolving} is a new modular design in the Android permission model, which separates the management of system permissions from custom permissions declared by untrusted third-party applications. It introduces backwards compatible naming conventions for custom permissions to systematically eliminate and prevent the loopholes of custom permissions. 

The mainstream approach for enhancing the Android permission mechanism is to identify over-declared permissions requested by an app \cite{checking2014gorla,aspg2014wang,autocog2014qu,whyper2013pandita}, and recommend appropriate permissions for an app \cite{skewness2016huang,automatic2019liu}. TERMINATOR \cite{temporal2018sadeghi} provides a safe, reliable, yet non-disruptive approach to protect mobile users against permission misuses.Bin Liu et al.\cite{liu2016follow} proposed a Personalized Privacy Assistant (PPA) for mobile applications, which can manage mobile permissions of mobile applications and predict the privacy settings that users want by asking some questions, and proposed a method to learn the privacy profile of permission settings. Bao \cite{automated2017bao} also propose two novel approaches to realize permission recommendation.

\section{Conclusion}

The mini program is a new mobile application format that runs inside a mobile app. Although these mini programs are taking over the traditional mobile OS and have become the way to do almost everything in China, there is little research on these mini programs, especially for their potential security and privacy issues. In this paper, we conducted a large-scale analysis of mini programs in different host apps for the first time. We have conducted empirical research on 9 currently popular host apps, revealing the security issues corresponding to the 6 types of potential security vulnerabilities we have discovered in the real world. We propose corresponding attack methods to analyze these potential weaknesses to exploit the discovered vulnerabilities. In addition, we also showed three real-life attacks on the mini program's permissions to prove that the revealed vulnerabilities may cause serious consequences in real-world use. Following the practice of responsible disclosure, we have also reported newly discovered vulnerabilities to relevant security platforms, among which the more serious vulnerabilities obtained the CVE number. Lastly, we put forward a series of suggestions for the future deployment of mini programs to protect users' privacy.



\end{document}